# Data augmented turbulence modeling for three-dimensional separation flows

Chongyang Yan (闫重阳) [a], Yufei Zhang (张宇飞) [a,†], Haixin Chen (陈海昕) [a]

[a] *School of Aerospace Engineering, Tsinghua University, Beijing 100084, China*

**Abstract:**

Field inversion and machine learning are implemented in this study to describe three-dimensional (3-D) separation flow around an axisymmetric hill and augment the Spart–Allmaras (SA) model. The discrete adjoint method is used to solve the field inversion problem, and an artificial neural network is used as the machine learning model. A validation process for field inversion is proposed to adjust the hyperparameters and obtain a physically acceptable solution. The field inversion result shows that the non-equilibrium turbulence effects in the boundary layer upstream of the mean separation line and in the separating shear layer dominate the flow structure in the 3-D separating flow, which agrees with prior physical knowledge. However, the effect of turbulence anisotropy on the mean flow appears to be limited. Two approaches are proposed and implemented in the machine learning stage to overcome the problem of sample imbalance while reducing the computational cost during training. The results are all satisfactory, which proves the effectiveness of the proposed approaches.

[†] Associate professor, corresponding author; Email: zhangyufei@tsinghua.edu.cn

---

## 1. Introduction

Accurate simulation of turbulent flows using computational fluid dynamics (CFD) is one of the critical problems in various engineering applications, including internal flows, such as compressor and turbine flows, and external flows, such as flow over a wing or fuselage. Typically, there are three approaches for turbulence modeling, namely, Reynolds-averaged-Navier-Stokes (RANS) modeling, large eddy simulation (LES), and direct numerical simulation (DNS). Although DNS and LES are scale-resolving and high-fidelity [1], the computational cost is too high for daily applications for design purposes. Consequently, RANS modeling remains the most dependable approach for engineering applications and is expected to continue to do so for the foreseeable future. [1]

However, RANS models do not make correct predictions in some complicated cases, including separated flow [2][3], secondary corner flow [4][5], and high streamline curvature [6][7]. Among them, describing separation flow over a smoothly curved three-dimensional hill is one of the most challenging problems for RANS modeling. Unlike separation from a sharp edge, in flow over a 3-D hill, both the separation and reattachment locations are sensitive to the development and skew of the boundary layer, which is affected by a smoothly varying adverse pressure gradient [8][8]. Simpson et al. conducted several experiments on different 3-D hill configurations using the laser Doppler velocimetry (LDV) technique [10][11], and the results are applied as test cases for many subsequent turbulent modeling studies, including RANS and LES studies. Although LES simulations usually yield satisfactory results compared with experiments [8][12], most RANS models make poor predictions of the mean flow characteristics [8][12].

The poor performance of RANS models in this case might be attributed to two reasons. First, the non-equilibrium turbulence in the separated shear layer on the leeward side of the hill might not be accurately described by the turbulence model. Non-equilibrium turbulence has been found in a series of 2-D flow cases, including periodic hills [5] and iced airfoils [13]. Non-equilibrium turbulence may be present in 3-D separated shear layers. Moreover, LES simulations have indicated that the production of turbulence kinetic energy increases intensely upstream of the separation zone due to the unsteady separation process at that location [8]. Current RANS models may not describe this behavior. Second, studies have shown that turbulence over a 3-D hill could be locally anisotropic [14], in which case a linear eddy viscosity model could not describe this behavior. For instance, it has been shown that linear RANS models could not describe the increasing the Reynolds normal stress at the separating shear layer observed in experiments [15]. The purpose of the present study is to construct an improved turbulence model that reflects these two effects and makes satisfactory predictions.

Given high-fidelity observations of data from experiments or DNS, the traditional approach to the development of new turbulence models usually includes manually constructing correction terms and manually adjusting the parameters. In recent years, data-driven approaches, such as machine learning and uncertainty quantification, have been incorporated into turbulence modeling [16][17]. For example, given complete high-fidelity data from the field, a machine learning model can be trained to predict the discrepancy between the Reynolds stresses obtained from RANS computations and those in DNS simulations at each grid point, where local flow quantities can be used as input features [18][19]. However, this approach is not suitable when the data is limited or indirect [20][21], for example containing flow quantities in only a specific region, or being integrated quantities such as the lift coefficient. Instead, a new paradigm, named field inversion and machine learning (FIML), has been proposed for limited observations [22][23]. In the FIML approach, the RANS equation is usually modified by a spatially varying correction factor. Field inversion is used to infer the value of the correction factor based on maximum a posteriori (MAP) inference given limited data, and a machine learning model is used to predict the correction factor using local flow features. This framework has been successfully applied to a series of cases, including flows around an airfoil at large angles [24], in turbomachinery [25], in transitions over boundary layers [22], and over 2-D and 3-D bumps [15].

Field inversion has been often interpreted in the framework of uncertainty inference [16]. It is assumed that the model-form uncertainty of the current RANS model comes from the correction factor $\beta$, and $\beta$ follows an a priori distribution $p(\beta)$. Given data $D$, the posterior distribution of $\beta$ is written based on the Bayesian formula:

$$p(\beta|D) \propto p(D|\beta)p(\beta) \qquad (1)$$

Field inversion implements MAP inference for $\beta$ by solving an optimization problem. Alternatively, the inference problem has been solved by Markov chain Monte Carlo (MCMC) methods [26][27] or ensemble Kalman filtering (EnKF) [28][29] to obtain more detailed information on the probability distribution.

There has been much research on the application of machine learning in turbulence modeling. Some of the most important issues have included the construction of input features and machine learning models. The input features have usually been built using a tensor basis and invariants according to tensor representation theory [31] or manually constructed quantities by physical intuition [30]. The selection and evaluation of the features have been discussed in detail by Xiao et al. [19] and Yin et al. [30] Various machine learning algorithms have been used for turbulence modeling. The most commonly used models have been artificial neural networks [24][25] and random forests [32]. Apart from these deterministic models, Parish et al. and Duraisamy et al. have used a Gaussian process [23] to obtain the output with its uncertainty, which is an advantage of probabilistic models. Other effective machine learning models have included sparse symbolic regression by Schmelzer [33] and the gene expression programming (GEP) method by Zhao et al [34], which could be expected to provide a symbolic turbulence model.

FIML has been challenged by the potential multiplicity of solutions in the field inversion stage. Gradient optimization algorithms have usually been used to apply field inversion as an optimization problem, and the inversion process can be easily trapped in a local optimum. Consequently, a regularization term has been used often to obtain physically reasonable solutions [35]. However, as shown in this paper, the hyperparameters of the regularization, together with other settings of field inversion, can greatly impact the final solution. More criteria are needed to validate and evaluate the solutions from field inversion, including data-driven and physically intuitive criteria, to obtain physically acceptable and interpretable solutions. The present study proposes an improved application of field inversion to incorporate manual intervention with physical knowledge and utilize limited data to full advantage. Another challenge has arisen from the imbalance of the samples for machine learning. The trivial samples with a correction factor that equals 1 (no correction) have accounted for most of the samples [24][25], and the problem is more severe in 3-D cases. In the present study, several approaches are tried and compared to address this problem.

In this paper, the improved FIML framework is implemented for data-driven augmentation of the Spart–Allmaras (SA) model to locate the defects of the present model for 3-D separated hill flows and improve the prediction accuracy. The rest of this paper is organized as follows. In section 2, the mathematical method of field inversion and machine learning in the present work is introduced in detail. In section 3, a validation case of fully developed incompressible duct flow is introduced to prove the effectiveness of the current framework. Next, the FIML framework is used for the 3-D hill flow case, and the results are presented in detail. Finally, section 4 concludes the paper.

## 2. Method

*2.1 Field inversion with the discrete adjoint method*

In this paper, the SA model is modified by two multipliers that reflect the non-equilibrium turbulent effect and the anisotropy of the Reynolds stress. The original SA model proposed by Spalart and Allmaras [36] is a one-equation model that is used to solve for the eddy viscosity $\widetilde{v}_t$:

$$\frac{D\tilde{v}}{Dt} = P - D + \frac{1}{\sigma}\left[\nabla \cdot \left((v + \tilde{v})\nabla\tilde{v}\right) + C_{b2}(\nabla\tilde{v})^2\right] \quad (2)$$

where $v$ represents the molecular viscosity. The eddy viscosity $v_t$ is computed by:

$$v_t = \tilde{v}f_{v1}, \quad f_{v1} = \frac{\chi^3}{C_{v1}^3 + \chi^3}, \quad \chi = \frac{\tilde{v}}{v} \quad (3)$$

where $P$ and $D$ are the production and destruction terms, respectively:

$$P = C_{b1}\tilde{S}\tilde{v}, \quad D = C_{w1}f_w\left[\frac{\tilde{v}}{d}\right]$$

$$\tilde{S} = (1 - f_{t2})\|S\| + \frac{\tilde{v}}{\kappa^2 d^2}[(1 - f_{t2})f_{v2} + f_{t2}]$$

$$f_w = g\left(\frac{1 + C_{w3}^6}{g^6 + C_{w3}^6}\right)^{\frac{1}{6}}$$

$$g = r + C_{w2}(r^6 - r)$$

$$r = \min\left(\frac{\tilde{v}}{\kappa^2 d^2 \hat{S}}, 10.0\right)$$

$$\hat{S} = \max\left(\|S\| + \frac{\tilde{v}f_{v2}}{\kappa^2 d^2}, 0\right)$$

$$f_{v2} = 1 - \frac{\chi}{1 + f_{v1}\chi}$$

$$f_{t2} = C_{t3}e^{-C_{t4}\chi^2} \quad (4)$$

Here, $S$ is the mean strain rate tensor, $d$ is the wall distance, $\kappa = 0.41$, $\sigma = 2/3$, and $C_{b1}, C_{b2}, C_{w1}, C_{w2}, C_{w3}, C_{v1}, C_{t3}$ and $C_{t4}$ are model constants.

To leverage the non-equilibrium effect in the SA model, a spatially varying correction factor $\beta(x)$ is multiplied with the product term in the $\tilde{v}_t$ equation:

$$\frac{D\tilde{v}}{Dt} = \beta(x) \cdot P - D + \frac{1}{\sigma}\left[\nabla \cdot \left((v + \tilde{v})\nabla\tilde{v}\right) + C_{b2}(\nabla\tilde{v})^2\right] \quad (5)$$

This is the same as the correction factor that has been used in previous FIML studies [24]. In addition, it is similar to the modifications of Rumsey [5] and Li [13] in which the destruction term of $\omega$ has been modified by a multiplier.

Although the inclusion of $\beta(x)$ greatly enhances the expression capability of the turbulence model, the model does not describe the anisotropy of Reynolds stress because of the Boussinesq assumption. As an extension, according to the non-linear eddy viscosity assumption, the Reynolds stress is expressed as a higher order polynomial about $S$ and the mean rotation rate tensor $\Omega$. According to tensor function representation theory, the polynomial can always be written as a combination of ten integrity basis components [31]. In this paper, a quadratic basis component $(\Omega S - S\Omega)$ is chosen and included in the constitutive relation of the SA model to maintain simplicity:

$$\tau_{ij} = \overline{\tau_{ij}} - \eta(x)\big[O_{ik}\overline{\tau_{jk}} - O_{kj}\overline{\tau_{ik}}\big]$$
$$\overline{\tau_{ij}} = 2\nu_t\left(S_{ij} - \frac{1}{3}u_{kk}\right), O_{ik} = \frac{\partial_k u_i - \partial_i u_k}{\sqrt{\partial_n u_m \partial_n u_m}} \quad (6)$$

where $\overline{\tau_{ij}}$ is the Reynolds stress calculated by the linear eddy viscosity constitutive relation, $O_{ik}$ is the dimensionless rotation rate tensor $\Omega$, and $\eta(x)$ is a spatially varying factor. In fact, when $\eta = 0.3$, the model reduces to the SA-QCR2000 model [37]. This is another important reason for choosing the quadratic tensor basis.

In general, $\boldsymbol{\beta}$ and $\boldsymbol{\eta}$ are random fields that obey a joint normal prior distribution. The mean and variance of the prior distribution are usually determined by physical knowledge. After the data $\boldsymbol{D}$ are given, the posterior distribution of $\boldsymbol{\beta}$ and $\boldsymbol{\eta}$ can be inferred according to the Bayesian formula. Assuming that $\boldsymbol{D}$ also obeys a joint normal distribution, the optimization object of field inversion can be written as:

$$J = \sum_i (d_i - h_i(\boldsymbol{\beta}, \boldsymbol{\eta}))^2 + \lambda \sum_j (\beta_j - 1)^2 + (\eta_j - 1)^2 \tag{7}$$

where $d_i$ represents the data at the i-th point, $h_i$ represents the quantity of interest at the i-th point computed by the RANS solver with given $\boldsymbol{\beta}$ and $\boldsymbol{\eta}$, and $\beta_j$ represents the correction factor at the j-th grid point. Each $d_i$ is assumed to be independently and identically distributed as are the $\beta_j$. The first term corresponds to the likelihood of the given $\boldsymbol{\beta}$ and $\boldsymbol{\eta}$, and the second term corresponds to the prior distribution of $\boldsymbol{\beta}$ and $\boldsymbol{\eta}$, which can be regarded as a regularization term. $\lambda$ is a parameter indicating the confidence of the data relative to the prior turbulence model.

The optimization problem is solved with the gradient-based nonlinear programming solver SNOPT [38], and the discrete adjoint method [39][40] is used to compute the gradients. An important advantage of the adjoint method is that the computational cost is independent of the number of design variables, which makes it suitable for the optimization problem of field inversion where the number of design variables equals that the number of grid cells, $\sim 10^6$ in this paper.

The process of the discrete adjoint method is briefly introduced as follows. If the RANS equations can be expressed as $R(\boldsymbol{x}, \boldsymbol{w}) = 0$, where $\boldsymbol{x}$ represents the design variables and $\boldsymbol{w}$ represents the flow quantities, and the object function is expressed as $J(\boldsymbol{x}, \boldsymbol{w})$, then the discrete adjoint method solves for the gradient of $J$ relative to $\boldsymbol{x}$:

$$\boldsymbol{G} = \frac{dJ}{dx} = \frac{\partial J}{\partial x} + \frac{\partial J}{\partial w} \frac{\partial w}{\partial x} \tag{8}$$

It is difficult to directly compute $\frac{\partial w}{\partial x}$. Instead, we can obtain $\boldsymbol{G}$ by the following

two steps:

(1) Compute $\frac{\partial \boldsymbol{R}^T}{\partial \boldsymbol{w}}$ and $\frac{\partial J}{\partial \boldsymbol{w}}$ and solve the discrete adjoint equation to obtain the adjoint variable $\boldsymbol{\varphi}$:

$$\left[\frac{\partial \boldsymbol{R}}{\partial \boldsymbol{w}}\right]^T \boldsymbol{\varphi} = -\left[\frac{\partial J}{\partial \boldsymbol{w}}\right]^T \tag{9}$$

(2) Compute $\frac{\partial J}{\partial \boldsymbol{x}}$ and $\frac{\partial \boldsymbol{R}}{\partial \boldsymbol{x}}$ and obtain $\boldsymbol{G}$:

$$\boldsymbol{G} = \frac{\partial J}{\partial \boldsymbol{x}} + \boldsymbol{\varphi}^T \frac{\partial \boldsymbol{R}}{\partial \boldsymbol{x}} \tag{10}$$

In this study, we modify the open-source CFD code ADFlow [41] for the RANS computation and discrete adjoint computation, in which the third party library Tapanade [42] is used for automatic differentiation.

*2.2 Machine learning*

In the FIML framework, a machine learning model is used to build the functional correlation locally between $\boldsymbol{\beta}, \boldsymbol{\eta}$ inferred from field inversion and the local flow features. After the model is trained to predict $\boldsymbol{\beta}, \boldsymbol{\eta}$ precisely, it is expected to have learned some generalizable knowledge on turbulence modeling. We choose an artificial neural network (ANN) as the machine learning model. Notably, we have tested other algorithms, such as the random forest algorithm, and their performance has been comparable to that of the ANN. We finally choose the ANN because it is the fastest in our framework.

We use several input features, mostly from reference [24]. The features are manually selected basic local flow quantities, including the ratio of the eddy viscosity to the molecular viscosity $\chi = \tilde{v}/v$, the production and destruction terms of $\tilde{v}$, the norm of the mean flow strain and rotation rate, and the norm of Reynolds stress $\tau$:

$$\left\{\chi, P, \frac{P}{D}, S, \frac{S}{\Omega}, \tau, f_d'\right\}$$

The feature $f_d'$ is initially proposed for detached eddy simulations [43] and modified in reference [24]:

$$f_d' = 1 - tanh(r_d^{0.5}) \tag{11}$$

Proper normalization of the input features is necessary for the generalization of the machine learning model. In the present work, we use physical scales combined from $(\tilde{v} + v)$ and wall distance $d$ to normalize the dimensional features:

$$\bar{S} = \frac{d^2}{\tilde{v}+v} S, \bar{\Omega} = \frac{d^2}{\tilde{v}+v} \Omega$$
$$\bar{P} = \frac{d^2}{(\tilde{v}+v)^2} P, \bar{D} = \frac{d^2}{(\tilde{v}+v)^2} D \tag{12}$$

The network contains three hidden layers, the sizes of which are 32, 16, and 8. The activation function is ReLU. Batch normalization is used at each hidden layer. The overall structure of the network is shown in Fig. 1. The open-source package PyTorch [44] is used to train the ANN models.

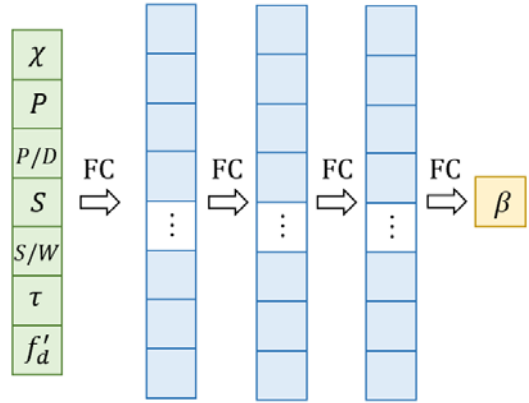

Fig. 1 The overall structure of the ANN

Imbalance of samples has been often encountered in FIML problems. Other studies have shown that the nature of the incremental correction determines that the non-trivial region, where the correction factor is not equal to 1.0, usually occupies a concentrated and narrow area [24][25]. Although the problem can be neglected by training with the full dataset (and the performance is satisfactory), the time required for training could still be reduced significantly by some specific treatment. One approach is to reserve only the samples inside a prescribed area containing the non-trivial region and discard the others. Another approach is to down sample from the trivial samples and implement a weighted regression. The two approaches are performed and compared in this study.

*2.3 Overall framework*

The key objective of FIML is to make generalizable modifications to existing turbulent models. In addition to the effective training of machine learning models, reasonable solutions from field inversion are also critical and should reveal the underlying physical mechanism. In practice, we find that the hyperparameters and other settings of field inversion can affect the results significantly, and they should be determined carefully. The following are the possible hyperparameters:

(1) The specified model-form uncertainties where the correction factors are added. Each correction factor corresponds to a specific physical impact.

(2) The data used in the object function. Not all the data can be used directly in the object function. Since the solution space of the modified RANS equation is limited [16], fitting too much data might overdetermine the problem. Besides, some data from observations might be non-quantitative, for example, photographs of surface streamlines.

(3) The options for the construction of the regularization term. There are other choices in addition to the independent identically distributed prior distribution, and physical knowledge is leveraged herein. It should be also noted that some knowledge is hard to be encoded directly into the expression of the regularization term.

(4) Other hyperparameters. For example, when heterogeneous data are used in the object function, the relative weights for different data types should be determined. Another parameter is the number of iterations for field inversion. Sometimes the sufficient convergence of the objective function might lead to overfitting, and an early stop is needed to obtain physical realizable solutions.

We propose an improved approach for the application of field inversion to obtain reasonable physical solutions by using a validation process to adjust the hyperparameters. As discussed above, some data from observation and physical knowledge might be excluded from the object function. We seek to utilize such information to full advantage by using them to evaluate the solution from a specific set of hyperparameters. The best hyperparameters are determined according to the validation performance, and the final solution is expected to reveal the underlying physical mechanism. The overall framework of the FIML framework is shown in Fig. 2. The training and validation of the machine learning model on the dataset from field inversion are performed without the CFD solver, which corresponds to the "offline" process; while during testing, the machine learning model is dynamically coupled with the CFD solver until convergence, which make the process "online".

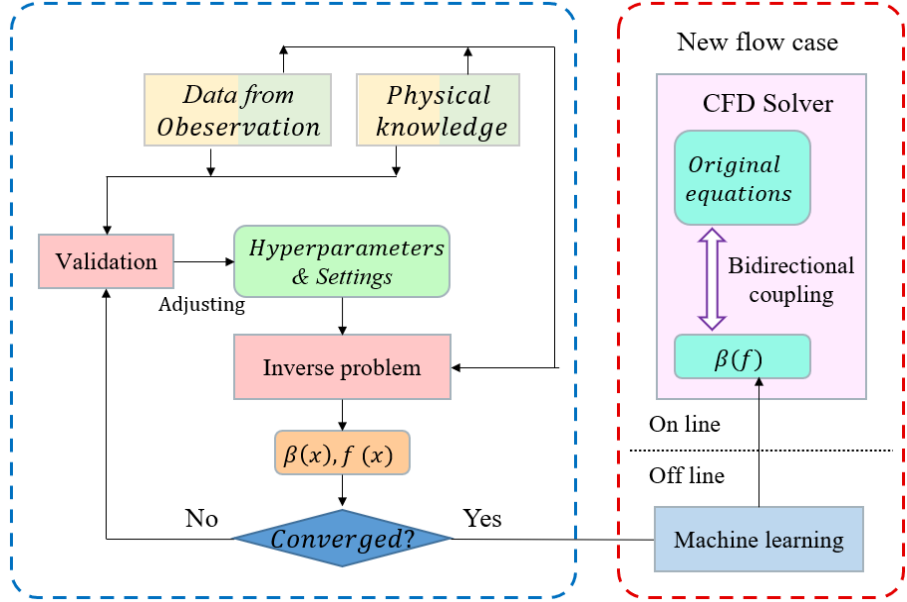

Fig. 2 The overall framework of the FIML framework

## 3. Results

*3.1 Fully developed square duct flow*

3.1.1 Case description

The FIML framework is first applied to the fully developed incompressible square duct flow problem to validate the effectiveness of the current framework. The square duct flow is a typical flow problem of turbulent anisotropy, which is characterized by the corner secondary flow [45]. The linear eddy viscosity model cannot predict the transverse flow in such corners. Consequently, the case has been used for the calibration and validation of many non-linear eddy viscosity models [37]. The DNS results from Huser et al. [46] are used as data from observation in this study. The Reynolds number based on the side length $D$ of the duct and mean velocity is $Re_D = 10320$. The computation domain is shown in Fig. 3. The length-width ratio L/D=6, and only 1/4 of the domain is reserved, considering symmetry. A symmetry plane boundary condition is applied on the two internal sections, periodic conditions are applied for the inlet and outlet sections, and no-slip wall conditions are applied on the wall surfaces. As the ADFlow code used in this study is used for compressible flows, the incompressible flow is approximated by the computation with a low Mach number (~0.1).

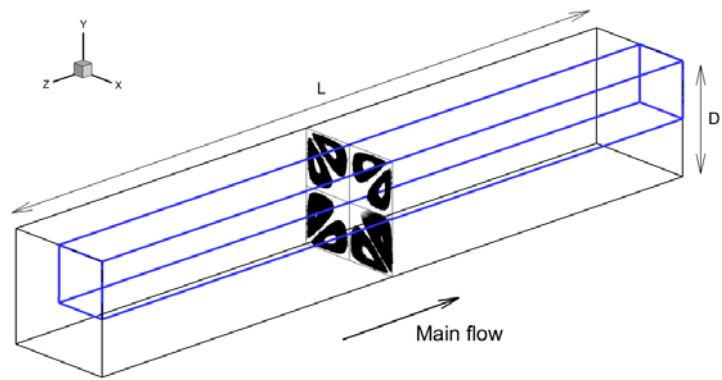

Fig. 3 The computation domain of the square duct flow

The mass flow rate at the inlet is prescribed to ensure a certain Reynolds number based on bulk velocity. Consequently, a uniform volume force term is added to the whole field to drive the flow, and the CFD solver automatically adjusts the magnitude of the volume force to match the prescribed mass flow rate. Since the flow problem is physically constrained, the augmented adjoint method [47] is used for field inversion.

### 3.1.2 Field inversion

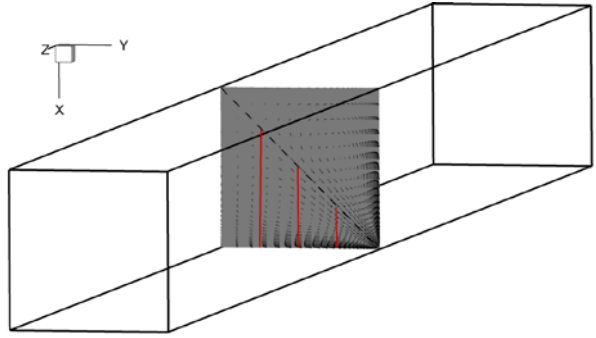

Fig. 4 The three vertical lines (red) along which the data are given

In this case, the limitation of the SA model comes from its inability to depict turbulent anisotropy. Only the factor $\eta$ (introduced in section 2.1) is added to make a non-linear constructive correlation. As shown in Fig. 4, the velocity profiles along three vertical lines starting from different spanwise locations of the wall from DNS are used as the data for field inversion. The factor $\eta$ is used as a constant in the whole field for simplicity, and the regularization term is omitted in this case. The objective function can be written as:

$$J = \sum_i \left( u_i - u_i^h(\eta) \right)^2 \tag{13}$$

The original $\eta$ was set to 0.09. As shown in Fig. 5, the field inversion converges within 6 steps. The convergence of the squared error of velocity is shown in Fig. 5(a), and the error is reduced to 12% of the original value. The velocity profiles from the original and inversed SA models compared with the DNS results are shown in Fig. 6. The transverse velocity predicted by the original SA model is 0, and the velocity profiles from the inversed SA model are relatively close to those of the DNS. As shown in Fig. 5(b), the factor $\eta$ converges to 0.284, which is similar to the corresponding parameter 0.3 in the SA-QCR2000 model [37]. As the form of the model is the same as that of the SA-QCR2000 model, the data-driven calibration of the model parameters and the manually calibrated parameters agree, which proves the effectiveness of the FIML framework used in this study.

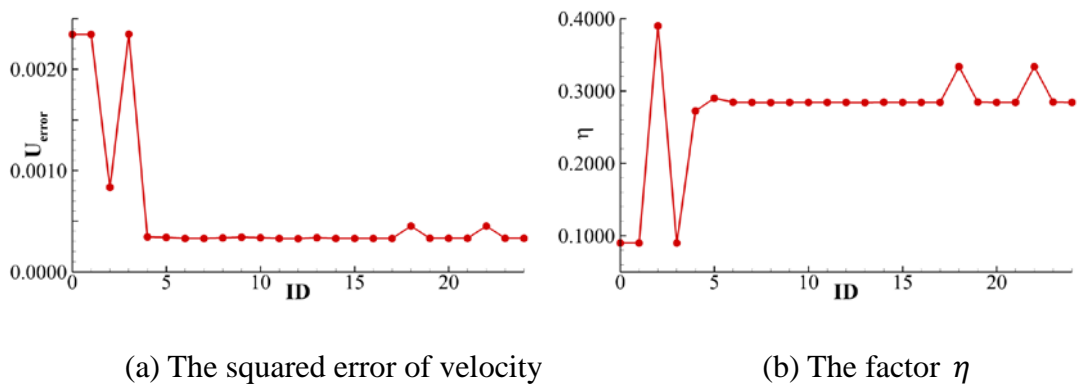

(a) The squared error of velocity    (b) The factor $\eta$

Fig. 5 The time records of the convergence of the squared errors of the velocity and the factor $\eta$.

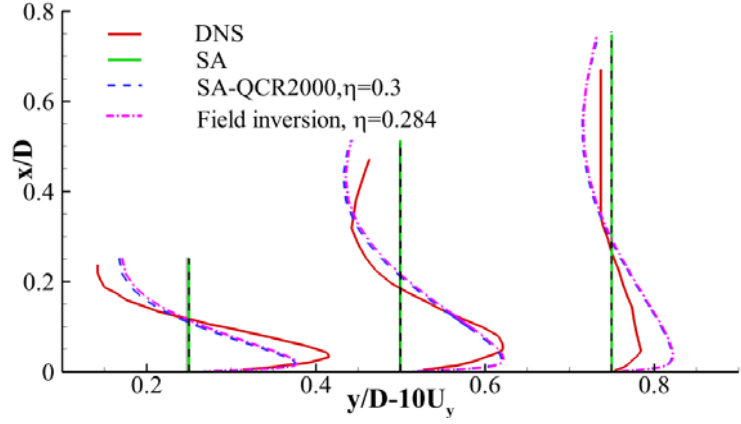

Fig. 6 Comparison of the velocity profiles from the original and inversed SA models with the results from the DNS and SA-CR2000 model.

*3.2 Three-dimensional hill flow case*

3.2.1 Case description

The flow over an axisymmetric 3-D hill is a case of complicated flow with three-dimensional separation over a smoothly varying curved surface. The geometry of the configuration and experiment data are obtained from a series of experiments on 3-D hills conducted by Simpson et al. [10][11]. The geometry of the configuration used for field inversion in this paper is axisymmetric, and the generatrix equation is:

$$\frac{y(r)}{H} = -\frac{1}{6.04844}[J_0(\Lambda)I_0(\Lambda(r/a)) - I_0(\Lambda)J_0(\Lambda(r/a))] \quad (14)$$

Here, $\Lambda = 3.1962$, $H = 0.078m$ is the height of the hill, $a = 2H$ is the radius of the hill, $J_0$ is the Bessel function of the first kind, and $I_0$ is the modified Bessel function of the first kind. Fig. 7 shows the geometry of the configuration of the 3-D hill and the computation domain for CFD. Only half of the domain is reserved because of symmetry.

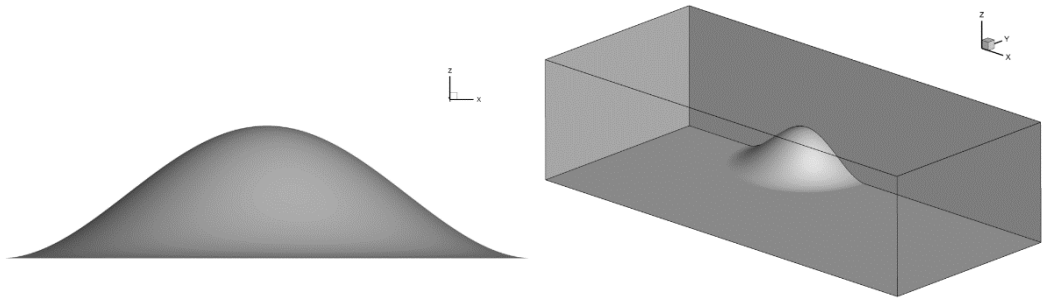

Fig. 7 The geometry of the configuration of the 3-D hill and the computation domain

The section of the grid for CFD computation on the symmetry plane is shown in Fig. 8. The grid dimension is $165 \times 69 \times 93$. Local refinement is implemented around the 3-D hill both in the streamwise and spanwise directions. The first layer of the grid in the boundary layer on the wall meets $y^+ < 1$, with a growth rate along the normal direction of the wall of 1.15.

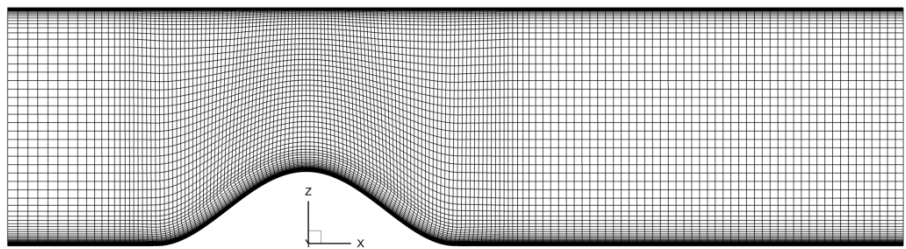

Fig. 8 The section of the computation grid for the 3-D hill on the symmetry plane

Subsonic inlet boundary conditions are applied at the inlet, with the velocity, density and eddy viscosity prescribed. A subsonic outlet condition with pressure prescribed is applied at the outlet. The symmetry plane condition is applied on the symmetry plane, and the no-slip wall condition is applied on the rest of the surfaces. To provide more information about the inlet condition for the duct, Simpson et al. conducted another experiment without the 3-D hill and measured the velocity profile at the location of *x*=0 (the peak of the hill), which is shown in Fig. 9. To make the inflow condition as similar as possible to the experiments, one approach by Wang et al. [8] has been to implement another RANS computation without the 3-D hill in a sufficiently long duct and find the section at which the velocity profile is the most similar to that shown in Fig. 9. Since the inlet section of the computation domain is at *x*=-4*H*, the flow conditions 4*H* upstream of the found section are used as the inlet conditions for the 3-D hill simulation. This approach is employed in the present work. The thickness of the inlet boundary layer is $\delta = 0.5H$, and the Reynolds number based on the hill height and mean velocity is $Re_H = 130000$.

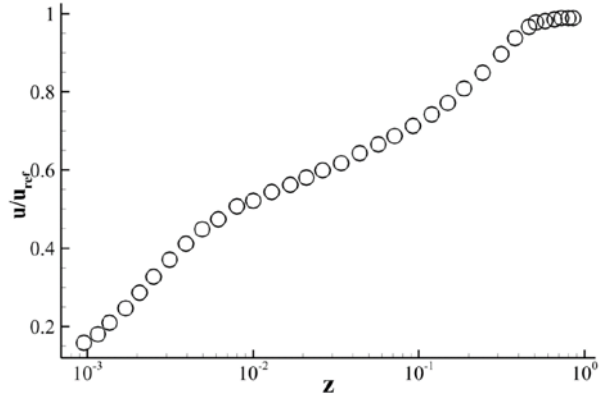

Fig. 9 The velocity profile at *x*=0 in the experiment without the 3-D hill

Many numeric studies have been conducted on the same geometry. Wang et al. [8] have tested several RANS models that all predicted an excessively large separation region compared to experiments, while LES computations [8] have usually rendered a separation region similar to that of experiments. García-Villalba et al. [8] have performed an elaborate analysis of their LES results. As shown in Fig. 10(a)(b), they have found an area with high turbulence upstream of the time-averaged separation region in both the LES and the experiment. As shown in Fig. 10(c), a region with high production of turbulent kinetic energy (TKE) upstream of the separation line leeward on the hill was also observed in LES [48]. They have speculated that the separation location varied instantaneously in the range of $x/H = 0.3{\sim}0.9$, and the production of turbulence in this area originated from the unsteady separation process. In fact, the overprediction of the separation region has been found in many 2-D flows, such as those of periodic hills [5] and iced airfoils [13]. The non-equilibrium turbulence in the separating shear layer has been proven to be the critical consideration needed for precise RANS predictions [49]. The ratio of the production to dissipation of the turbulent kinetic energy is usually near 1.0 for equilibrium turbulence, however this value could be much larger than 1.0 in separating shear layer [50]. Ordinary RANS models have not reflected this behavior, so the turbulence mixing in the separated shear layer has been underpredicted, and the reattachment of the separating flow has been delayed [13]. We suspect that the separation region is also strongly affected by non-equilibrium turbulence in 3-D separating flows.

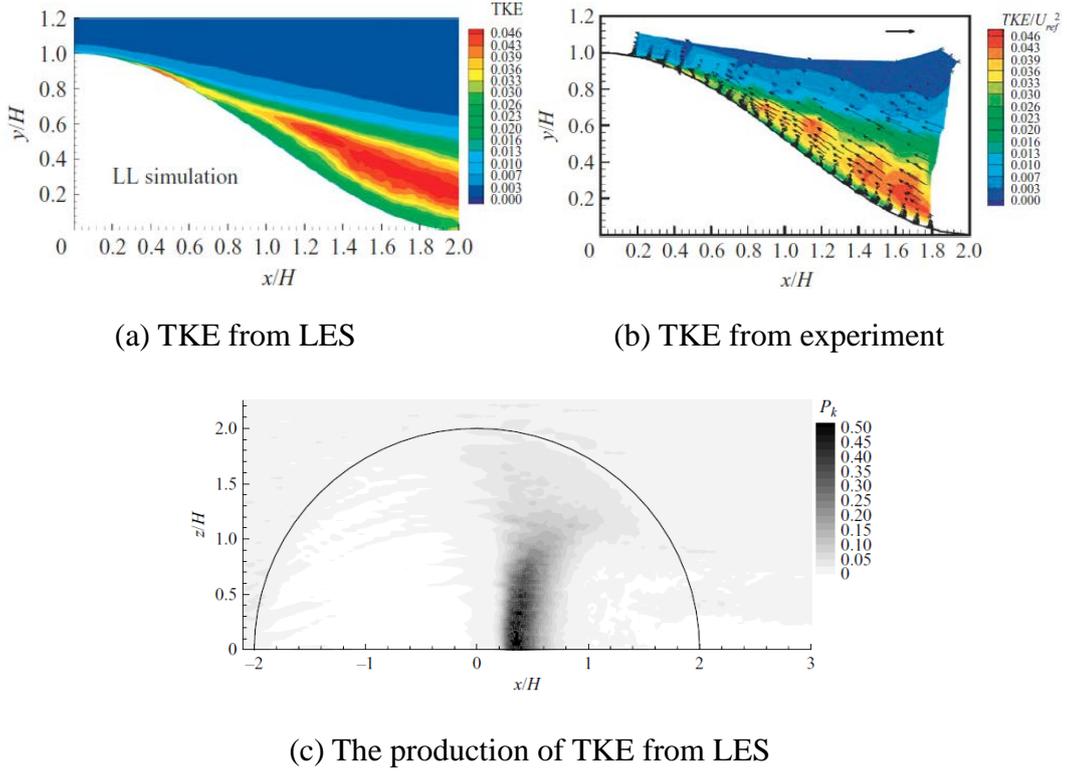

(a) TKE from LES  (b) TKE from experiment

(c) The production of TKE from LES

Fig. 10 The turbulence kinetic energy and its production term from LES and

experiment [8]

3.2.2 Field inversion

As discussed above, the correction factor of the production term of eddy viscosity $\beta$ should be added as an uncertain parameter. In addition, since turbulence anisotropy has been observed in experiments for the 3-D Fundamental Aeronautics Investigates The Hill (FAITH hill) in a separating shear layer [15], the quadratic Reynolds stress term is also introduced in the present work with the uncertain parameter $\eta$. The number of optimization variables reaches $2.2 \times 10^6$ when the cells in the whole field are used for inference, which generates unnecessary cost of memory and time from the optimization algorithm. Consequently, the candidate region for correction is restricted to a relatively small cuboid around the 3-D hill according to the physical knowledge that the limitations of the SA model are mostly present in the separating shear layer. Specifically, the correction area is in the region of $0 < x/H < 3.69$, $0 < y/H < 2.0$, and $0 < z/H < 1.0$, and the number of optimization variables was reduced to $2.4 \times 10^5$.

Heterogeneous data are obtained from experiments [11], and the profiles of the streamwise and spanwise velocities at different spanwise locations at $x/H = 3.69$, together with the pressure coefficient distribution along the surface of the 3-D hill in the streamwise direction at $z/H = 1.0$, are used in the object function of field inversion. A simple regularization term from an independent identically distributed Gaussian prior is added. The object function of field inversion is written as:

$$J = \sum_i \left(u_i - u_i^h(\boldsymbol{\beta}, \boldsymbol{\eta})\right)^2 + \xi \sum_j \left(C_{p_j} - C_{p_j}^h(\boldsymbol{\beta}, \boldsymbol{\eta})\right)^2 + \zeta \sum_k [(\beta_k - 1.0)^2 + (\eta_k - 0.3)^2] \quad (15)$$

Notably, other experimental data are not included directly in the object function, including photographs of the streamlines on the hill surface and the vector clouds on the symmetric plane of the flow domain and at the section downstream at $x/H = 3.69$. These data, together with other prior physical knowledge discussed above, make very important contributions to the validation of the results from field inversion.

Table 1 The best combination of the hyperparameters of field inversion for the 3-D hill

| Hyperparameter | Value |
| --- | --- |
| Data from observation | $u, v$ at $x/H = 3.69$, $C_p$ distribution at $z/H = 1.0$ |
| Uncertain parameters | $\boldsymbol{\beta}, \boldsymbol{\eta}$ |
| Regularization term | $\sum_k [(\beta_k - 1.0)^2 + (\eta_k - 0.3)^2]$ |
| Weights of losses | $\xi = 0.1, \zeta = 1 \times 10^{-5}$ |
| Number of iterations | 6 |

The best combination of the hyperparameters of field inversion is shown in Table 1. The $u, v$ profiles at $x/H = 3.69$ from field inversion are compared with the results from the original SA model and experiment in Fig. 11. The fuller streamwise velocity profiles at the locations away from the symmetry plane from the original SA model indicate a much narrower wake region than observed in the experiment. As discussed below, the difference originates from the inability of the original SA model to capture the streamwise vortex which causes momentum transportation in the wake region[8]. For the solution from field inversion, the accuracy of the prediction is improved in most locations, although some differences are present. Fig. 12 shows the pressure coefficient distribution from field inversion at two spanwise locations compared with the results from the original SA model and experiment. The differences between the results from the original SA model and experiment are mainly present in the leeward part of the hill. The pressure recovery is much slower in the original SA model than in the experiment downstream of the separation location, while the pressure recovery is significantly improved in the solution from field inversion.

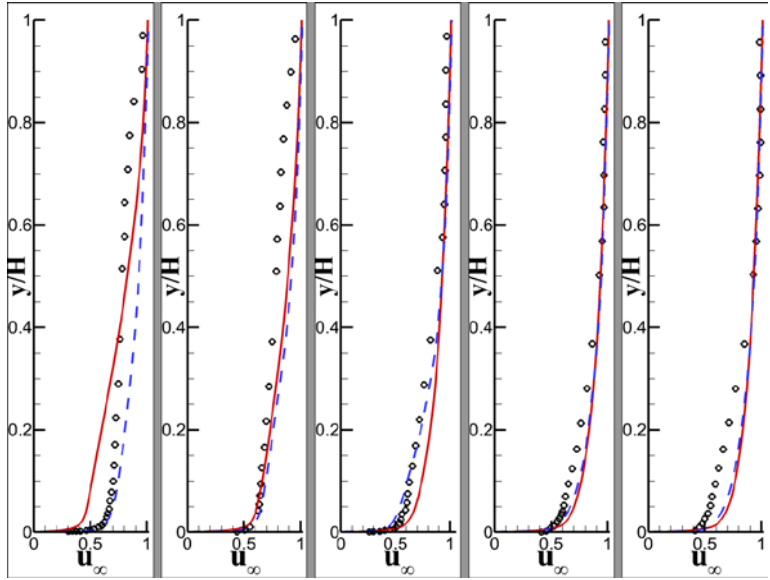

(a) Streamwise

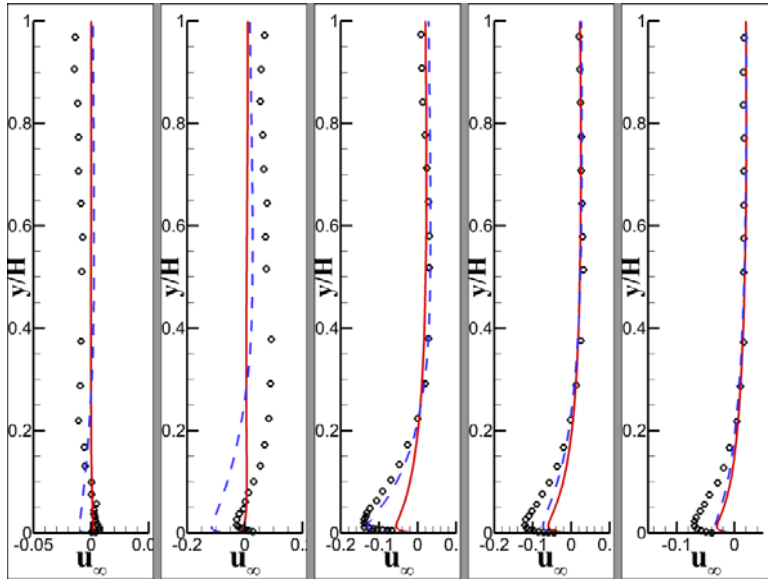

(b) Spanwise

Fig. 11 The velocity profiles at 5 spanwise locations at *x*/*H*=3.69 from field inversion compared with the original SA model and experiment. The 5 spanwise locations from left to right are: $z = 0, z/H = -0.33, z/H = -0.81, z/H = -1.30, z/H = -1.7$.

Black circles: experimental results, red solid lines: the original SA model, blue dashed lines: field inversion

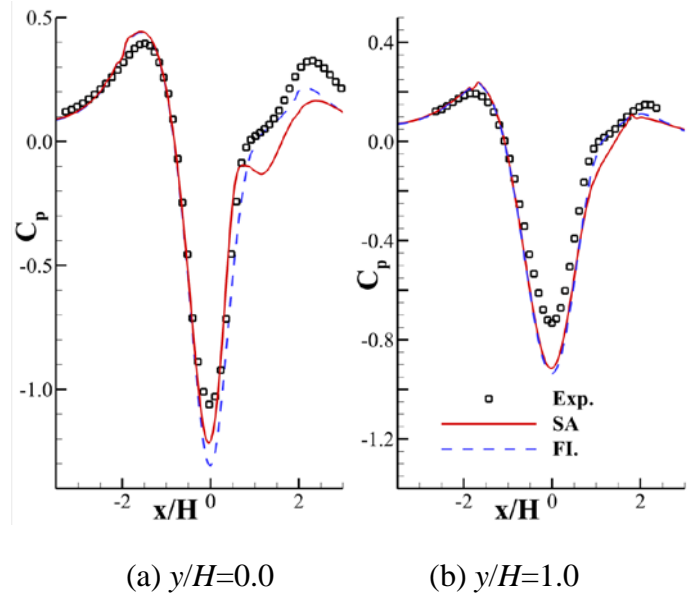

(a) $y/H=0.0$      (b) $y/H=1.0$

Fig. 12 Distributions of the pressure coefficient at two spanwise locations on the surface of a hill: field inversion (FI), the original SA and experimental (Exp) results.

Fig. 13 compares the results for the flow field on the symmetry plane of the flow domain from field inversion, the original SA model and experiments. The original SA model notably overpredicts the separation region on the leeward side of the hill. The length and height of the recirculation region exceed those from the experiment, which cover a rather narrow and relatively short region. We find that in the field inversion results, the recirculation region is significantly more narrow, similar to the experimental result, and satisfactory.

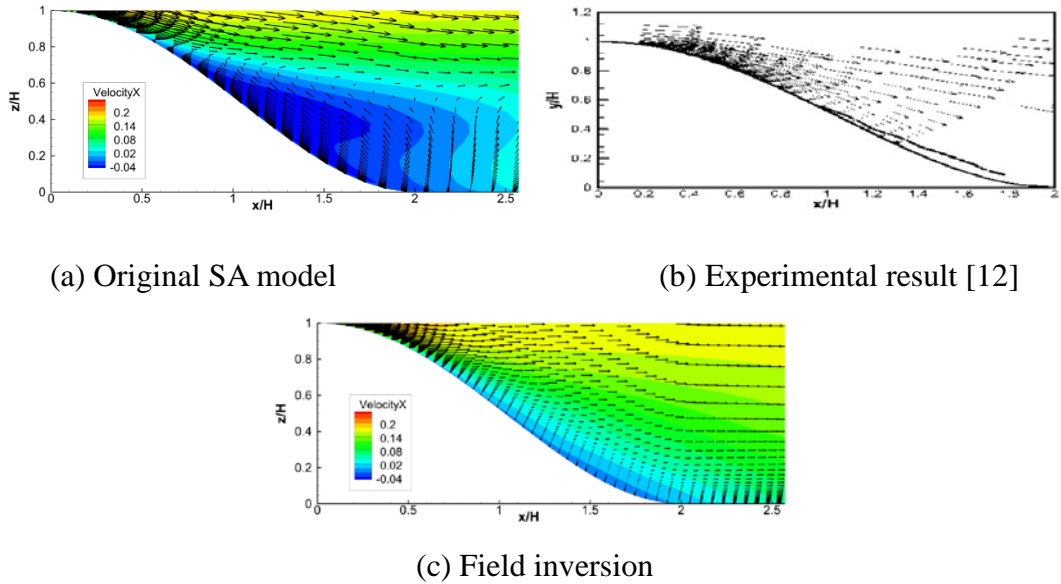

(a) Original SA model            (b) Experimental result [12]

(c) Field inversion

Fig. 13 Comparison of the flow fields on the symmetry plane of the flow domain from field inversion, the original SA and experiment

Fig. 14 shows the surface streamlines on the hill. The separation location on the symmetry plane is $x/H$ =0.5 for the original SA model, while the experimental location is $x/H$=1.0. In contrast, the separation and reattachment locations on the symmetry plane from field inversion are both very similar to the experiment results. In addition, compared to the experimental results, the surface vortex at the focal point from the original SA model is much stronger, and that from field inversion is significantly weaker. This indicates that the field inversion method gave improved predictions of the 3-D separation characteristics.

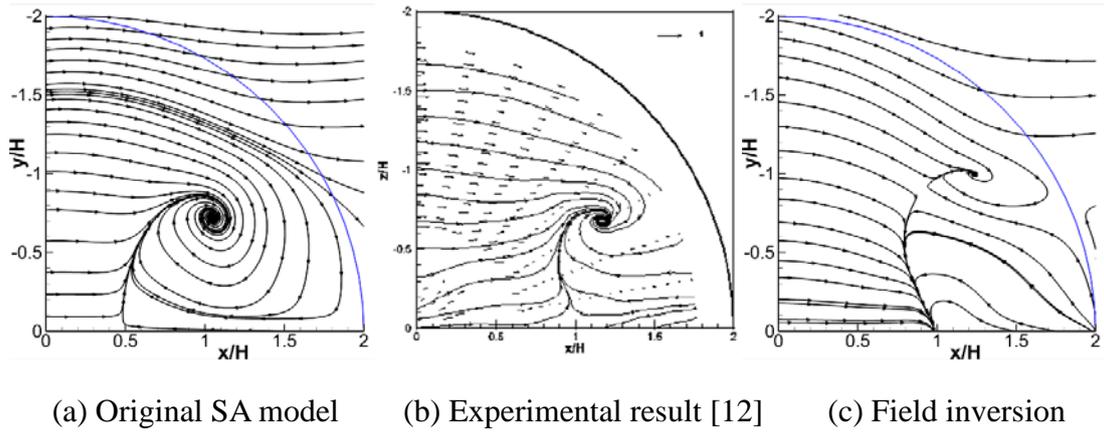

(a) Original SA model    (b) Experimental result [12]    (c) Field inversion

Fig. 14 The surface streamlines on the hill from field inversion, the original SA method and experimental results

Fig. 15 shows the flow field on the section downstream of the hill at $x/H = 3.69$. A strong artificial streamwise vortex is observed near the symmetry plane from the original SA model. The spatial streamlines suggest that this vortex is shed from the excessively strong focal point on the hill surface shown in Fig. 14(a). In contrast, this vortex is fully eliminated by field inversion in Fig. 15(c), which is due to the weakening of the artificial vortex at the focal point. Additionally, a main vortex rotating in the opposite direction is observed in the experiment, but it is almost eliminated by the original SA model. A previous study has demonstrated that this vortex originates from the impingement of the flow on the bottom wall downstream of the recirculation region at $x/H = 2.0$ [12]. Fig. 15(c) shows that the strong streamwise vortex in the experiment is accurately described by field inversion, which implies that the impingement of the flow is predicted accurately by the inverse solution. Finally, it can be inferred that the streamwise vortex in the weak region dominates the velocity transportation in that region, which explains the difference between the velocity profiles in Fig. 11 for the original SA model and field inversion.

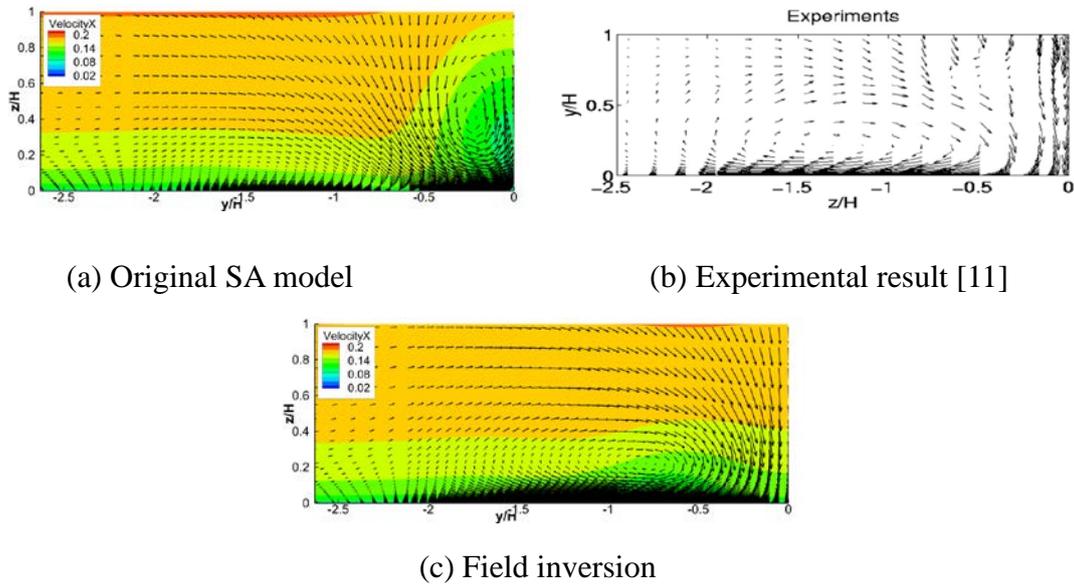

(a) Original SA model　　　　　　　　(b) Experimental result [11]

(c) Field inversion

Fig. 15 The flow field on the section downstream of the hill at $x/H = 3.69$ from field inversion, the original SA and experimental results

Fig. 16 shows the contours of the correction factors $\beta$ and $\eta$ obtained by field inversion; this region is concentrated in a narrow area. The production of turbulence is enhanced in the boundary layer upstream of the mean separation location leeward of the hill, which is consistent with the LES and experimental observations in Fig. 10. Enhanced production is also found in the starting part of the separating shear layer. This confirms the speculation about non-equivalent turbulence in 3-D separating flows. The turbulence anisotropy is enhanced in the separating shear layer, which is consistent with the observation in the 3-D FAITH hill experiment.

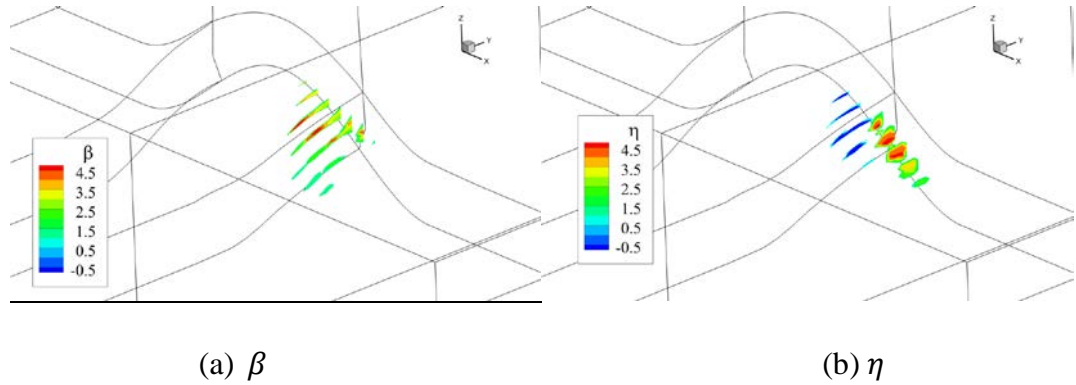

(a) $\beta$        (b) $\eta$

Fig. 16 The contour of the correction factors obtained by field inversion. The regions where $0.95 < \beta, \eta < 1.05$ are masked.

In general, the solution from field inversion agrees well with most of the data from the experiment, whether directly added into the object function. The corrected area reveals the underlying physical mechanism, and it is consistent with our physical knowledge. Some solutions from field inversion with suboptimal hyperparameters are shown here to illustrate the importance of the validation process. Three combinations of hyperparameters are selected and listed in Table 2; only one parameter differs from the optimal set in each case.

Table 2 Three suboptimal combinations of the hyperparameters of field inversion

| Case number | Different hyperparameter | Value |
| --- | --- | --- |
| 1 | Number of iterations | 30 |
| 2 | Data from observation | Only $u, v$ at $x/H = 3.69$ |
| 3 | Uncertain parameters | Only $\beta$ |

The quantities of interest (QoIs) of the three cases are shown in Fig. 17 and Fig. 18. The data appear to agree well with given experiment observations, which indicates the full convergence of the optimization problems. However, further comparison shows that these results perform worse in the validation stage. Fig. 19 shows the surface streamlines on the 3-D hill for the three cases. Compared to the experimental results in Fig. 14, the streamlines are excessively distorted for cases 1 and 2. Consequently, there is a large gap between the solutions from these cases and the ground truth. This indicates overfitting by the field inversion.

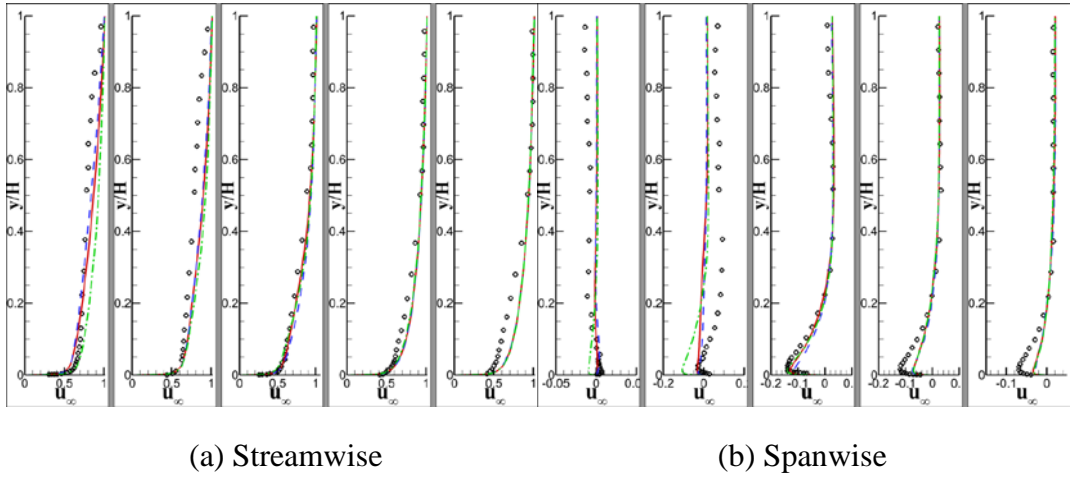

(a) Streamwise      (b) Spanwise

Fig. 17 The velocity profiles at 5 spanwise locations at $x/H$=3.69 from the three field inversion cases. Black circles: experiment, red solid lines: case 1, blue dashed lines: case 2, green dashed-dotted lines: case 3.

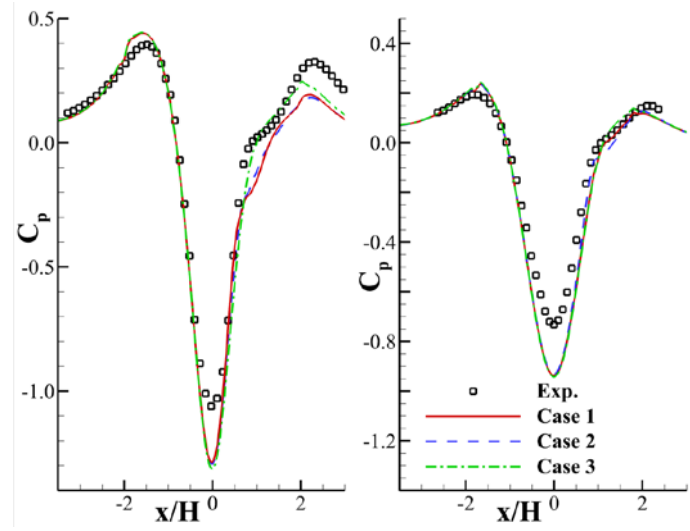

(a) *y/H*=0.0     (b) *y/H*=1.0

Fig. 18 The pressure coefficient distributions on the hill surface at two spanwise locations from the three field inversion cases

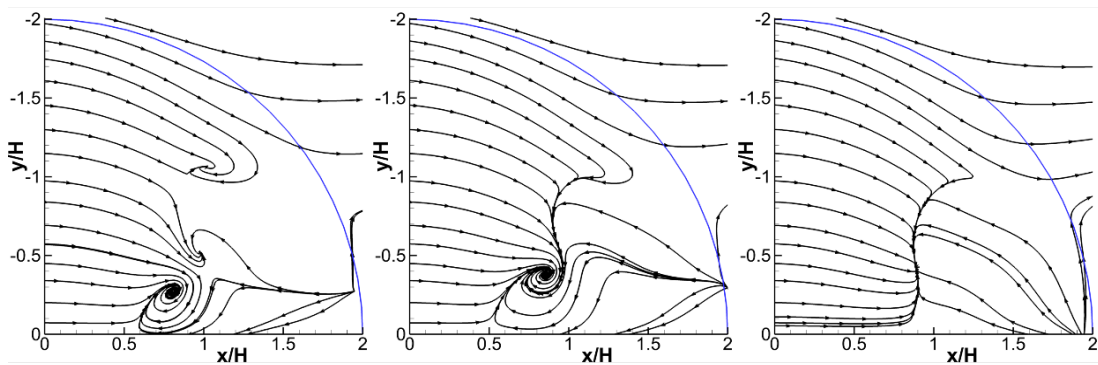

(a) Case 1     (b) Case 2     (c) Case 3

Fig. 19 The surface streamlines on the 3-D hill from the three field inversion cases

The contours of the correction factor $\beta$ for the three cases are shown in Fig. 20. The distributions of $\beta$ in cases 1 and 2 are indeed consistent with the physical knowledge of the non-equilibrium turbulence in the separating shear layer on the symmetry plane. However, the solutions are discarded because the predicted surface streamlines are inconsistent with those from experiments. Notably, the distribution of $\beta$ in case 3 is similar to the results in Fig. 16. The envelope of the corrected area is almost identical to that in Fig. 16, and the amplitude of correction is even lower, which shows more consistency with the previous model and suggests potentially better generalizability for machine learning. We can at least infer that for 3-D separation flow around an axisymmetric hill, the non-equilibrium turbulence effect has a decisive effect on the mean flow, while the impact of turbulence anisotropy is weak.

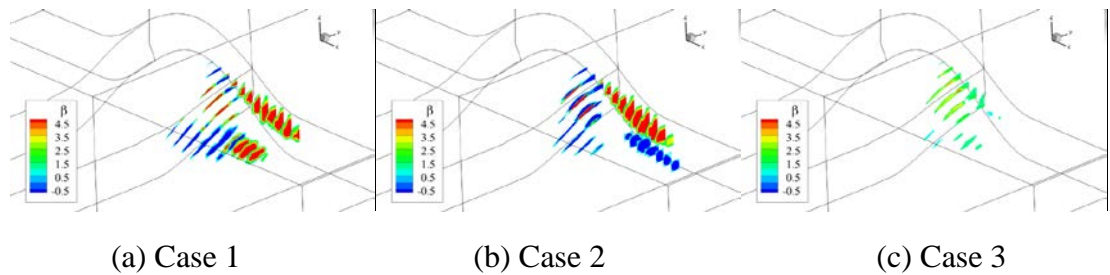

(a) Case 1　　　　　　　(b) Case 2　　　　　　　(c) Case 3

Fig. 20 The contours of the correction factor $\beta$ from the three field inversion cases

3.2.2 Machine learning

The results from field inversion are used to train a machine learning model to build a functional relationship between the local flow features and the correction factors. Two independent ANN models are trained for $\beta$ and $\eta$.

Since the ANN model will be integrated into the RANS solver and yield the correction factor at each grid point at each iteration, the ANN is used outside of the prescribed cuboid region of field inversion. To ensure that the ANN model makes no corrections in such a region, one intuitive approach is to gather the samples both inside and outside of the prescribed region and manually assign $\beta = \eta = 1.0$ for the samples outside of the prescribed region. This provides $2.18 \times 10^6$ samples for machine learning. Fig. 21 shows the histograms of the labels of the samples on a log scale. The trivial samples, for which $0.95 \leq \beta \leq 1.05$ or $0.95 \leq \eta \leq 1.05$, account for over 99.85% of the samples. This ratio is more exaggerated than in the two-dimensional cases.

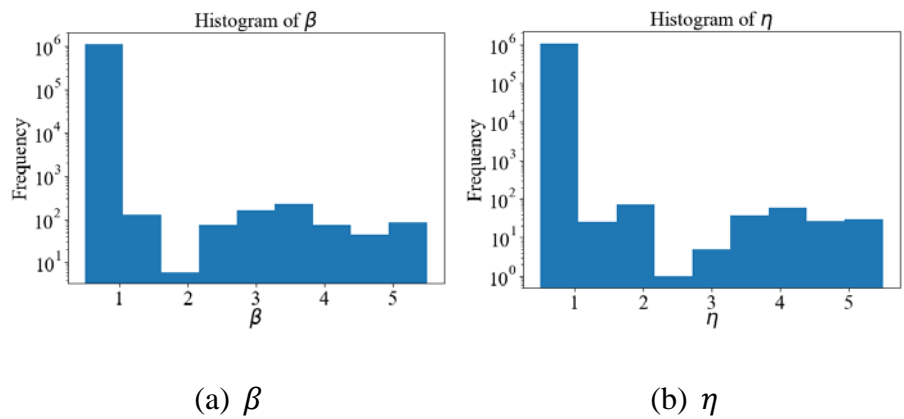

(a) $\beta$     (b) $\eta$

Fig. 21 Histograms of the labels of the samples from field inversion

It is possible to use all the samples for training a machine learning model, but this approach is costly in terms of memory and time. However, there are methods that can be used to train a machine learning model with sufficient accuracy and lower cost. One approach is to simply disregard the samples outside of the prescribed region of field inversion during training and directly assign the value 1.0 to the correction factors outside of the region while serving. This approach has been used in the prediction of eddy viscosity discrepancy by Yin et al. [50] Another approach is to down sample from trivial samples with a certain ratio $\alpha$ ($\alpha = 0.01$ in this paper) while reserving all the non-trivial samples. To offset the bias caused by the change in sample distribution, the weights of the losses from the trivial samples are set to $1/\alpha$, which leads to a weighted regression. This is equivalent to an ordinary regression after resampling on the sampled trivial samples with ratio $1/\alpha$. The underlying assumption is that the flow field outside of the prescribed region is smooth enough to ensure that after sampling, the distribution covers the corresponding feature space. This is consistent with our physical knowledge, i.e., the flow field changes greatly around the hill surface, while the opposite is true far from the hill. The number of samples can be further reduced by this approach because there are still many trivial samples in the prescribed region in field inversion.

Three approaches, namely, training with the full set of data (ANN1), a truncated set of data (ANN2) and sampled data (ANN3), are implemented in this paper, and the results are analyzed. The number of training epochs is 1000 for each ANN. The root mean squared error (RMSE) on the overall trivial and non-trivial samples for the three models are shown in Table 3. The offline errors on the samples are similar for the three approaches.

Fig. 22 shows the predictions of $\beta$ and $\eta$ using the offline features on several sections. The distributions of $\beta$ and $\eta$ are similar to the field inversion results except for the value of $\eta$ predicted by ANN3.

Table 3 The RMSEs for the three ANNs

| ANN | Number of training samples | RMSE of the trivial samples | RMSE of the non-trivial samples |
| --- | --- | --- | --- |
| ANN1 | $2.18 \times 10^6$ | 0.023 | 0.488 |
| ANN2 | $2.37 \times 10^5$ | 0.067 | 0.718 |
| ANN3 | $2.34 \times 10^4$ | 0.059 | 0.494 |

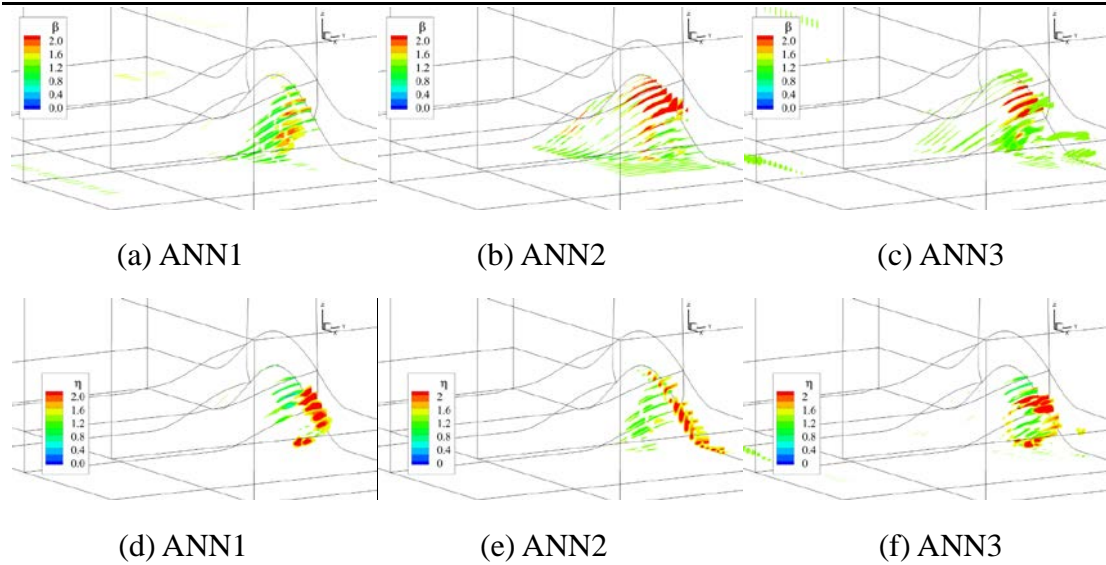

(a) ANN1  (b) ANN2  (c) ANN3

(d) ANN1  (e) ANN2  (f) ANN3

Fig. 22 The results for $\beta$ and $\eta$ predicted using the offline features on several sections by the three ANNs

The trained ANN models are integrated into the RANS solver to build an enhanced SA model. Bidirectional coupling is adopted during CFD computations, in which the ANN is called at each cell at each iteration to predict the correction factor for the turbulence equation with local flow features. The enhanced SA models coupled with the three ANN models are tested under the same inflow conditions as the field inversion. Fig. 23 shows the velocity profiles at $x/H = 3.69$, and Fig. 24 shows the pressure coefficient distribution predicted by the enhanced SA models. All the results are very similar to the solution from field inversion.

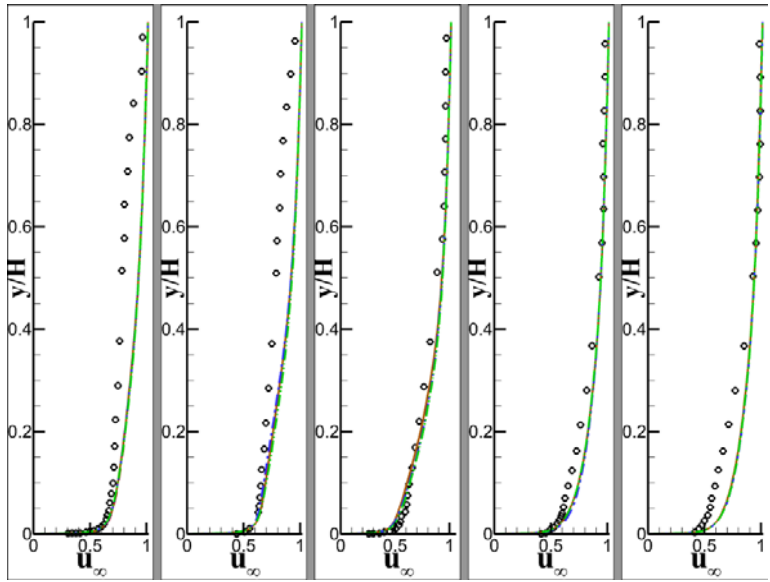

(a) Streamwise

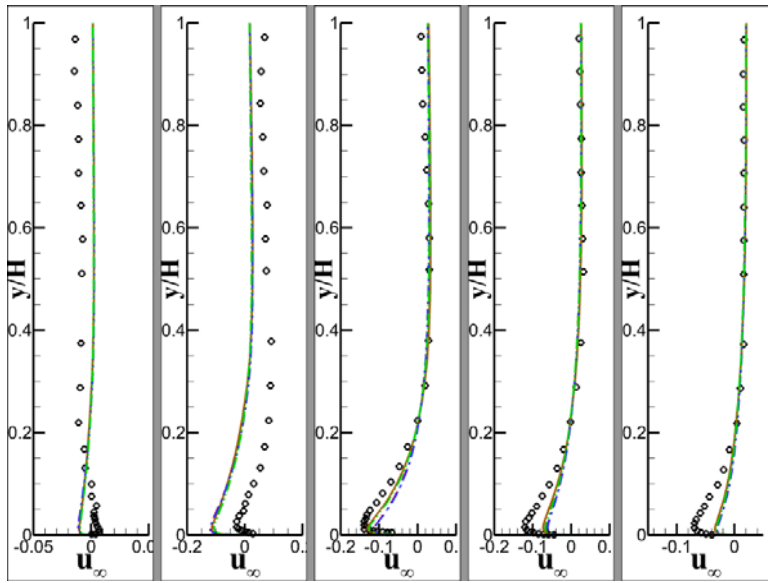

(b) Spanwise

Fig. 23 Comparison of the velocity profiles at $x/H = 3.69$ for the SA model coupled with the three ANNs with the results from field inversion. Black circles: experimental, brown solid lines: field inversion, red dashed lines: ANN1, blue dashed-dotted lines: ANN2, green dashed-dotted lines: ANN3.

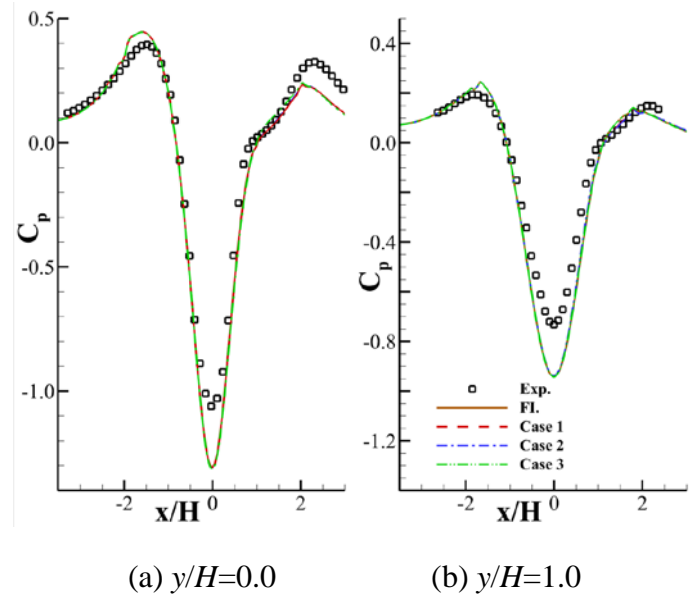

(a) *y/H*=0.0    (b) *y/H*=1.0

Fig. 24 Comparison of the pressure coefficient distributions predicted by the SA model coupled with the three ANNs with the results from field inversion

Fig. 25 shows the flow field on the symmetry plane of the duct, and Fig. 26 shows the streamlines on the hill surface. All the results are similar to the solution from field inversion; however, the separation point is slightly upstream on the symmetry plane. This shows the satisfactory performance obtained when the ANNs are used in a bidirectional coupling scenario. The results also indicate that the choice of the training approach has little influence on the final results, which proves the effectiveness of the training approaches.

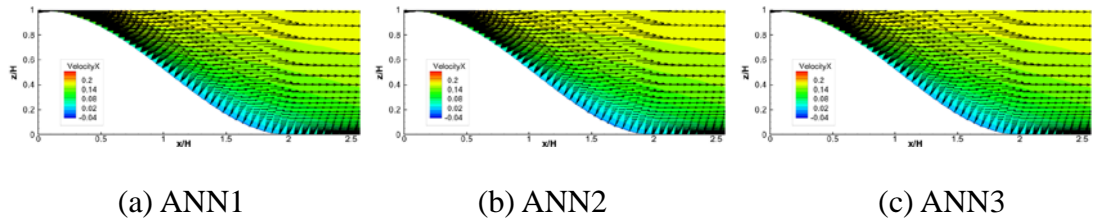

(a) ANN1　　　　　　　(b) ANN2　　　　　　　(c) ANN3

Fig. 25 The flow field on the symmetry plane of the duct for the enhanced SA models

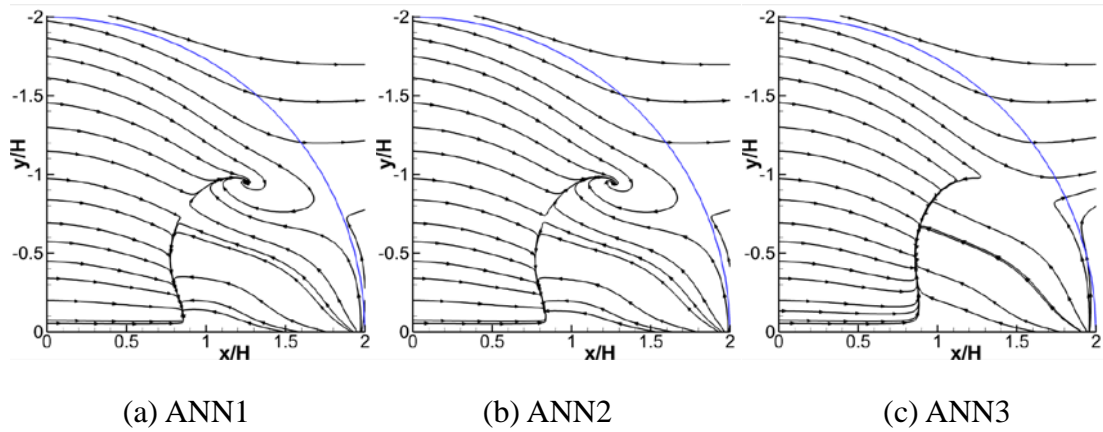

(a) ANN1　　　　　　　(b) ANN2　　　　　　　(c) ANN3

Fig. 26 The streamlines on the surface of the hill for the enhanced SA models

Fig. 27 shows the correction factors predicted by the integrated ANNs in converged solutions. Although the envelope of the corrected region is different from that of field inversion, the tendencies for increasing turbulence production upstream of the separation line and in the separation shear layer are accurately described. It can be inferred that the ANN models have learned the physical knowledge instead of simply remembering the corrections for specific flow features.

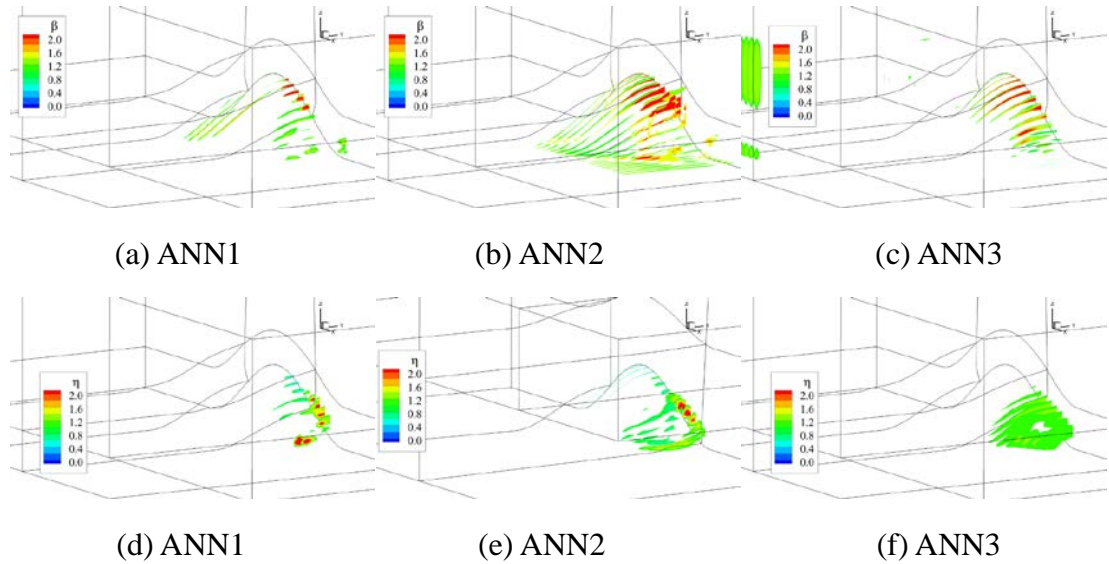

| (a) ANN1 | (b) ANN2 | (c) ANN3 |
| (d) ANN1 | (e) ANN2 | (f) ANN3 |

Fig. 27 The correction factors predicted by the integrated ANNs in converged solutions

The enhanced SA model is further tested on an unseen geometry: the 3D FAITH hill [51]. The Reynolds number based on the hill height is 500000, higher than that of the 3-D hill for field inversion (130000), which makes it a stringent test condition. The flow fields on the symmetry plane computed by the original SA and the enhanced SA models are compared with the experimental results in Fig. 28. The SA model overpredicts the range of the separation bubble, and the ANN-enhanced SA model shortens the separation region. Fig. 29 compares the velocity profiles on the symmetry plane. Both the SA and the enhanced SA models show some discrepancies with respect to the experimental results, while downstream of the separation region, the enhanced SA model makes better predictions than the enhanced SA model. The results show that the enhanced SA model can be generalized to a certain extent.

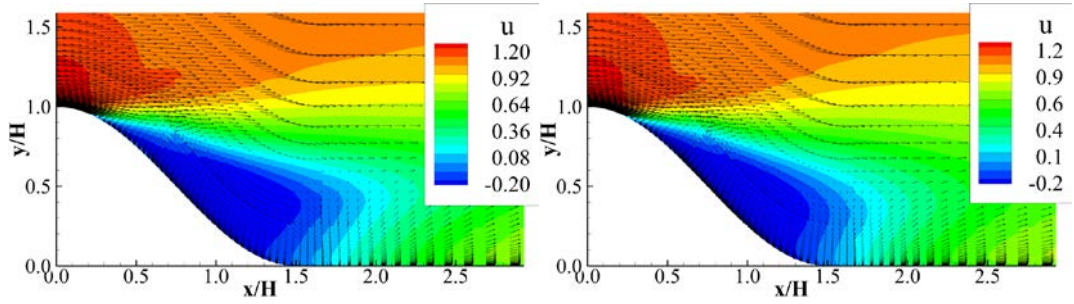

(a) The SA model  (b) The FIML enhanced SA model

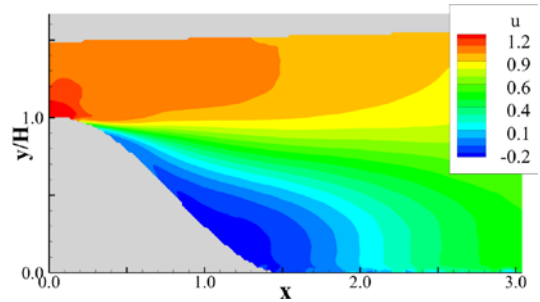

(c) Experiment

Fig. 28 Comparison of the flow fields on the symmetry plane for the FAITH hill obtained from field inversion, the original SA model and experimental results

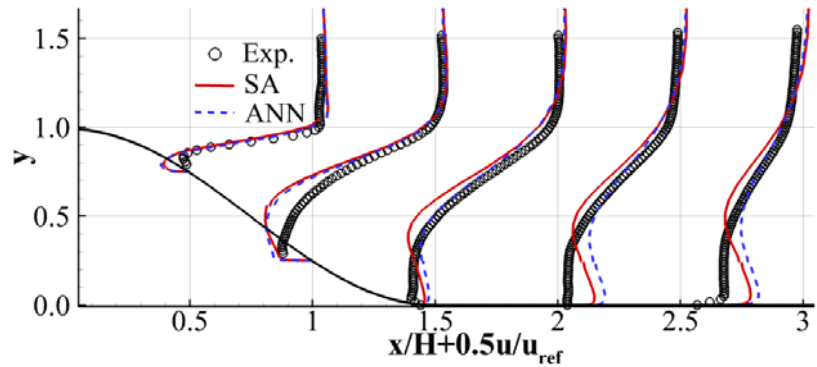

Fig. 29 Comparison of the velocity profiles on the symmetry plane for the FAITH hill from field inversion, the original SA model and experimental results

## 4. Conclusions

In this study, the SA model is augmented by the improved FIML framework to enhance the prediction of 3-D separation flows. The framework is first validated for a simple case of incompressible, fully developed duct flow. Next, field inversion is implemented on a 3-D axisymmetric hill case with complicated 3-D separation and wake structures. The trained machine learning models show satisfactory performance when integrated into the SA model in a bidirectional coupling scenario. Additional tests on an unseen geometry show that the enhanced SA model can be generalized.

Heterogeneous and limited data from experiments are used for field inversion. The solutions from field inversion are sensitive to the hyperparameters, and the validation process proposed in this paper proves to be critical to obtaining a physically reasonable solution. The corrections from the final solution indicate that the non-equilibrium turbulence effects in the boundary layer upstream of the mean separation location and inside the separating shear layer dominate the flow structures in a 3-D separation flow. This is consistent with our prior knowledge. Notably, the effect of Reynolds stress anisotropy on the mean flow appears to be limited.

Three machine learning models are trained with the full set of data, a truncated set of data and sampled data. The computational costs are lower when the truncated set of data and the sampled data are used instead of the full set. The results all show satisfactory performance, which proves the effectiveness of the two approaches. The correction area obtained from validating computations indicates that rather than simply remembering the specific mapping relationships, the models have learned generalizable knowledge about turbulence modeling.

## 5. Acknowledgments

This work was supported by the National Natural Science Foundation of China (grant nos. 92152301, 11872230, 92052203 and 91952302) and the Aeronautical Science Foundation of China (grant no. 2020Z006058002).

## 6. Conflict of interest statement

The authors have no conflicts to disclose.

## 7. Data availability statement

The data that supports the findings of this study are available within the article.